\documentclass[11pt]{iopart}

\newcommand{\bea}{\begin{eqnarray}}
\newcommand{\eea}{\end{eqnarray}}

\begin{document}

%%%%%%%%%%%%%%%%%%%%%%%%%%%%%%%%%%%%%%%%%%%%%%%%%%%%%%%%%%%%%%%
\title{Fully nonlinear and exact perturbations of the Friedmann world model: Non-flat background}
\author{Hyerim Noh}
\address{Korea Astronomy and Space Science Institute,
         Daejeon 305-348, Republic of Korea}
\ead{hr@kasi.ac.kr}
%\author{Hyerim Noh}
%\address{Korea Astronomy and Space Science Institute,
%         Daejeon 305-348, Republic of Korea}
%\ead{hr@kasi.re.kr}

%%%%%%%%%%%%%%%%%%%%%%%%%%%%%%%%%%%%%%%%%%%%%%%%%%%%%%%%%%%%%%%
%\date{\today}

%%%%%%%%%%%%%%%%%%%%%%%%%%%%%%%%%%%%%%%%%%%%%%%%%%%%%%%%%%%%%%%
\begin{abstract}

We extend the fully  non-linear and exact cosmological perturbation equations in a Friedmann background universe
to include the background curvature. The perturbation equations are
presented in a gauge ready form, so any temporal gauge
condition can be adopted freely depending on the problem to be solved.
%The background curvature term explicitly appears only in the energy and momentum constraint equations.
We consider the scalar, and vector perturbations without anisotropic stress.
As an application, we analyze the equations in the special case of irrotational zero-pressure fluid in the comoving gauge condition.
We also present the fully nonlinear formulation for a minimally coupled scalar field.

\end{abstract}

%%%%%%%%%%%%%%%%%%%%%%%%%%%%%%%%%%%%%%%%%%%%%%%%%%%%%%%%%%%%%%%
%\noindent \pacs{04.25.Nx, 98.80.Jk, 98.80.-k}

%%%%%%%%%%%%%%%%%%%%%%%%%%%%%%%%%%%%%%%%%%%%%%%%%%%%%%%%%%%%%%%

%%%%%%%%%%%%%%%%%%%%%%%%%%%%%%%%%%%%%%%%%%%%%%%%%%%%%%%%%%%%%%%
%\tableofcontents
%\maketitle

%%%%%%%%%%%%%%%%%%%%%%%%%%%%%%%%%%%%%%%%%%%%%%%%%%%%%%%%%%%%%%%
%
%  Introduction
%
%%%%%%%%%%%%%%%%%%%%%%%%%%%%%%%%%%%%%%%%%%%%%%%%%%%%%%%%%%%%%%%
\section{Introduction}
                                       \label{sec:Introduction}
The relativistic cosmological perturbation theory is the one of most important theoretical tools
to explain the cosmological observations.
Bardeen \cite{Bardeen-1988} presented a powerful cosmological perturbation formulation in which the
temporal gauge conditions can be applied freely depending on the
characters of the problem to solve \cite{Bardeen-1988}.
Bardeen's linear cosmological perturbation formulation was extended in \cite{Hwang-1991, Hwang-2001, Third-order-2005}
and to the second order \cite{NL-2004}.
In the recent work \cite{Hwang-2013}, we have extended Bardeen's formulation to the fully non-linear cosmological perturbations in the flat Friedmann universe. In \cite{Hwang-2013} the equations are presented in the gauge ready form, so the temporal gauge is not fixed. We assumed the flat Friedmann background universe.
We considered the scalar, and vetor perturbations of a fluid without anisotropic stress.
As is explained in \cite{Hwang-2013}, the transverse and tracefree tensor perturbations are not considered because of the
technical difficulties.
Instead, the tensor perturbation can always be handled to the nonlinear order perturbatively.

In this work, as one of the extensions we consider the fully non-linear perturbations in Friedmann background universe with
the general background curvature.

In Section 2 we summarize the metric and the energy momentum tensor convention and the gauge conditions.
In Section 3 we present the exact and fully nonlinear perturbation equations.
In Section 4, as an application we show the case in the comoving gauge.
We present the closed form of second-order differential equation for the perturbed energy density $\delta$.
Especially, the zero-pressure case was analyzed in detail.
In Section 5 we present the fully nonlinear perturbation equations for the minimally coupled scalar field
in a non-flat background model.
In Section 6 the discussion is given.
In the Appendix, the details useful for deriving the fully nonlinear perturbation equations are shown.

%%%%%%%%%%%%%%%%%%%%%%%%%%%%%%%%%%%%%%%%%%%%%%%%%%%%%%%%%%%%%%%
%
%
%
%%%%%%%%%%%%%%%%%%%%%%%%%%%%%%%%%%%%%%%%%%%%%%%%%%%%%%%%%%%%%%%
\section{Convention and gauge condition}
                                             \label{sec:convention}

In perturbed Friedmann universe with a background curvature the metric is given as
\bea
   & &
       ds^2 = - a^2 \left( 1 + 2 \alpha \right) d \eta^2
       - 2 a^2 \left( \beta_{,i} + B^{(v)}_i \right) d \eta d x^i
   \nonumber \\
   & & \qquad
       + a^2 \left[ \left( 1 + 2 \varphi \right) g^{(3)}_{ij}
       + 2 \gamma_{,i|j}
       + C^{(v)}_{i|j} + C^{(v)}_{j|i} + 2 C^{(t)}_{ij} \right]
          d x^i d x^j,
   \label{metric}
\eea where $a(\eta )$ is the cosmic scale factor. We assume $B^{(v)i}_{\;\;\;\;\;\;|i} \equiv 0 \equiv C^{(v)i}_{\;\;\;\;\;\;|i}$, and $C^{(t)i}_{\;\;\;\;\;i} = 0 = C^{(t)j}_{\;\;\;\;\;i|j}$ with indices of $B^{(v)}_i$, $C^{(v)}_i$ and $C^{(t)}_{ij}$ raised and lowered by $g^{(3)}_{ij}$ as the metric; $g^{(3)}_{ij}$ becomes $\delta_{ij}$ in a flat background; indices $(v)$ and $(t)$ indicate the vector- and tensor-type perturbations, respectively; a vertical var indicates the
covariant derivative based on the $g^{(3)}_{ij}$ as the metric tensor. Indices $a, b, \dots$ indicate the spacetime indices, and $i, j, \dots$ indicate the spatial ones.
The metric $g^{(3)}_{ij}$ is the background comoving three-space part of the Robertson-Walker metric, and is given as
\bea
  & & g^{(3)}_{ij}dx^i dx^j
     = {dr^2\over 1-{\overline K}r^2}
      + r^2 (d\theta^2 + \sin^2\theta d\phi^2 )
      \nonumber \\
  & & \qquad  = {1\over (1+{ {\overline K}\over 4} \bar r^2)^2} (dx^2 + dy^2 + dz^2 )
      \nonumber \\
  & & \qquad = \ d\bar \chi^2 + \Bigg[ {1\over \sqrt{\overline K}}\sin(\sqrt{\overline K}\bar\chi )\Bigg]^2
       (d\theta^2 + \sin^2\theta d\phi^2 ),
\eea
where
\bea
  & & r\equiv {\bar r\over 1+{\overline K\over 4} \bar r^2}, \quad
      \bar r\equiv \sqrt{ x^2 + y^2 + z^2}, \quad
            \bar\chi \equiv \int^r {dr\over \sqrt{1-\overline Kr^2}}.
\eea
The background curvature $\overline K$ is defined in equation (\ref{background-curvature}).

We decompose the spatial vector into longitudinal
and transverse parts as $B_i = \beta_{,i} + B^{(v)}_i$, and a
symmetric spatial tensor into longitudinal, trace, transverse, and
tracefree-transverse parts as $C_{ij} = \varphi g^{(3)}_{ij} +
\gamma_{,i|j} + {1 \over 2} (C^{(v)}_{i|j} + C^{(v)}_{j|i} ) +
C^{(t)}_{ij}$ \cite{York-1973}. All spatial indices
are raised and lowered by $g^{(3)}_{ij}$ as the metric. The transverse
part corresponds to the vector-type perturbation, and the tracefree-transverse
part corresponds to the tensor-type perturbation. The longitudinal
and trace parts correspond to the scalar-type perturbation.
%We can decompose the  perturbations order by order to
%nonlinear order.
In the homogeneous-isotropic background the three types of perturbations
decouple from each other, only to the linear order. In the nonlinear order we have couplings
among the scalar-, vector- and tensor-types of perturbations.
As in \cite{Hwang-2013}, here we consider only
the scalar- and vector-type perturbations.
%can be regarded as our main {\it assumption} in %this work. In Section \ref{sec:GW}, though, we will %consider the contribution of tensor-type perturbation %to the linear order.

%In considering the linear perturbation theory, Bardeen has suggested to take the spatial gauge condition
As the spatial gauge condition we choose
\bea
   & & \gamma \equiv 0 \equiv C^{(v)}_i.
   \label{spatial-gauge}
\eea
This is the only spatial gauge condition which does not leave the
gauge mode and allows the derivation of the fully nonlinear perturbation equations; see Section 2 of \cite{Hwang-2013}.
Then,  our metric becomes
 \bea
   & &
       ds^2 = - a^2 \left( 1 + 2 \alpha \right) d \eta^2
       - 2 a \chi_{i} d \eta d x^i
          + a^2 \left( 1 + 2 \varphi \right) g^{(3)}_{ij} d x^i d x^j,
   \label{metric-convention}
\eea
where
\bea
   & & \chi_i
       = a \left( \beta_{,i} + B^{(v)}_i \right)\equiv \chi_{,i} + \chi^{(v)}_i.
\eea
Considering an ideal fluid and choosing the energy frame we have the energy momentum tensor \cite{Ehlers-1993, Ellis-1971, Ellis-1973, Hawking-Ellis-1973}.
\bea
   & & \widetilde T_{ab} = \widetilde \mu \widetilde u_a \widetilde u_b
       + \widetilde p \left( \widetilde u_a \widetilde u_b + \widetilde g_{ab} \right) ,
 \label{Tab}
\eea where $\widetilde \mu$ and $\widetilde p$ are the energy density and the pressure, respectively, $\widetilde  u_a$ is the normalized fluid four-vector with $\widetilde u^a \widetilde u_a \equiv -1$;
%, and $\widetilde \pi_{ab}$ is the anisotropic stress with $\widetilde \pi_{ab} = \widetilde \pi_{ba}$, %$\widetilde \pi^a_a %\equiv 0$, and $\widetilde \pi_{ab} \widetilde u^b \equiv 0$
tildes indicate the covariant quantities.
%In this work we consider an ideal fluid with $\widetilde \pi_{ab} = 0$.
We define
\bea
  \fl \widetilde \mu \equiv \mu + \delta \mu \equiv \mu \left( 1 + \delta \right), \quad
       \widetilde p \equiv p + \delta p, \quad
       \widetilde u_i \equiv {a\over c} \widehat \gamma \widehat v_{i}, \quad
       \widehat \gamma \equiv  {1\over \sqrt{ 1-{\widehat v^k \widehat v_k \over c^2 (1+2\varphi )}}},
   \label{fluids}
\eea
where $\mu$ and $p$ are the background energy density and pressure, respectively; $\widehat\gamma$ is the Lorentz factor.

We decompose
\bea
   & & \widehat v_i \equiv - \widehat v_{,i} + \widehat v^{(v)}_i,
   \label{velocity}
\eea with $\widehat v^{(v)i}_{\;\;\;\;\;\;|i} \equiv 0$; $\widehat v_i$ and $\widehat v^{(v)}_i$ are raised and lowered by $g^{(3)}_{ij}$ as the metric.
The perturbed metric and fluid quantities
could have arbitrary amplitude.

Our fully nonlinear equations are arranged without fixing the temporal gauge conditions which we call the
``gauge ready" methods.
As Bardeen introduced \cite{Bardeen-1988}, the gauge ready method has the advantage
in the sense that any gauge conditions can be adopted  depending on the problem at hand.
We have several temporal gauge conditions available.
The fundamental gauge conditions are
the comoving gauge ($\widehat v=0$), the zero-shear gauge ($\chi =0$), the uniform-curvature gauge ($\varphi =0$),
the uniform expansion gauge ($\kappa = 0$), the uniform-density gauge ($\delta = 0$), and the
synchronous gauge ($\alpha = 0$); for $\kappa$, see later.
Except for the synchronous gauge, all above gauge conditions remove the gauge mode completely, and the
variables in those gauge conditions are gauge invariant.
The detailed gauge issues to the nonlinear order are explained
in Section 2 of \cite{Hwang-2013}.

%%%%%%%%%%%%%%%%%%%%%%%%%%%%%%%%%%%%%%%%%%%%%%%%%%%%%%%%%%%%%%%
\section{Exact and fully nonlinear perturbation equations in a gauge-ready form}
                                             \label{sec:equations}

Based on the ADM (Arnowitt-Deser-Misner) and the covariant formulation \cite{ADM}, \cite{Ehlers-1993}-\cite{Ellis-1973} we can derive the fully non-linear perturbation equations in a non-flat Friedmann background universe; the ADM and the covariant set of equations are presented in the Appendix A and C of \cite{Hwang-2013}, and the quantities useful for derivation are presented in the
Appendix.
An overdot indicates the covariant derivative based on t with $cdt\equiv ad\eta$.

\noindent
Definition of $\kappa$:
\bea
  \fl \kappa
       + 3 H \left( {1 \over {\cal N}} - 1 \right)
       + {1 \over {\cal N} (1 + 2 \varphi)}
       \left[ 3 \dot \varphi
       + {c \over a^2} \left( \chi^k_{\;\;|k}
       + {\chi^{k} \varphi_{,k} \over 1 + 2 \varphi} \right)
       \right]
       = 0.
   \label{eq1}
\eea

\noindent
ADM energy constraint:
\bea
  \fl - {3 \over 2} \left( H^2 - {8 \pi G \over 3c^2} \widetilde \mu  + {\overline K c^2\over a^2 (1+2\varphi )} -{\Lambda c^2 \over 3} \right)
       + H \kappa
       + {c^2 \Delta \varphi \over a^2 (1 + 2 \varphi)^2}
       \nonumber \\
  \fl \qquad    = {1 \over 6} \kappa^2
       - {4 \pi G\over c^2} \left( \widetilde \mu + \widetilde p \right)
       (\widehat\gamma^2-1)
              + {3 \over 2} {c^2\varphi^{|i} \varphi_{,i} \over a^2 (1 + 2 \varphi)^3}
              -{c^2\over 4}\overline{K}^i_j \overline{K}^j_i.
   \label{eq2}
\eea

\noindent
ADM momentum constraint:
\bea
   \fl {2 \over 3} \kappa_{,i}
        +{c\over a^2 {\cal N} (1+2\varphi )} \left[ {1\over 2}\left( \chi^{\; \; |k}_{i \; \; \; \; k}
          + \chi^k_{\; \; |ik} \right) -{1\over 3} \chi^k_{\; \; |ki}\right]
              + {8 \pi G\over c^4}  \left( \widetilde \mu + \widetilde p \right) a \widehat\gamma^2 \widehat v_{i}
              \nonumber \\
   \fl \qquad
   ={c\over a^2 {\cal N} (1+2\varphi )}
    \Bigg\{ \Big( {{\cal N}_{,j}\over {\cal N}} -{\varphi_{,j}\over 1+2\varphi }\Big)
    \Big[ {1\over 2} (\chi^j_{\; \; |i} +\chi_i^{\; \; |j} )
           -{1\over 3} \delta^i_i \chi^k_{\; \; |k} \Big]
           \nonumber \\
    \fl \qquad \quad
        -{\varphi^{|j}\over (1+2\varphi )^2 }
         \Big( \chi_i \varphi_{,j} +{1\over 3} \chi_j \varphi_{,i} \Big)
           + {{\cal N}\over 1+2\varphi } \Big[ {1\over {\cal N}}
          \Big( \chi^j \varphi_{,i} +\chi_i \varphi^{|j}
                -{2\over 3} \delta^j_i \chi^k \varphi_{,k} \Big) \Big]_{|j}
           \Bigg\}.
           \label{eq3}
       \eea

\noindent
Trace of ADM propagation:
\bea
   \fl - 3 \Bigg[ {1 \over {\cal N}} \dot H
       +H^2
       +{4\pi G\over 3 c^2} (\widetilde \mu + 3\widetilde p )
       -{\Lambda c^2\over 3} \Bigg]
       + {\dot\kappa\over \cal N}
       +2H\kappa
       +{c^2 {\cal N}^{|k}_{\; \; \; \; k}\over a^2{\cal N} (1+2\varphi )}
       \nonumber \\
   \fl   \qquad
       ={1\over 3}\kappa^2
        +{8\pi G\over c^2} (\widetilde\mu + \widetilde p ) (\widehat\gamma^2 -1 )
        -{c\over a^2 {\cal N} (1+2\varphi )}
        \Big( \chi^i \kappa_{,i} + c{\varphi^{|i}{\cal N}_{,i}\over 1+2\varphi } \Big)
         + c^2 \overline K^i_j \overline K^j_i.
          \label{eq4}
\eea

\noindent
Tracefree ADM propagation:
\bea
   \fl
       \left( {1\over {\cal N}}{\partial\over \partial t} + 3H -\kappa \right)
       \Bigg\{ {c\over a^2 {\cal N} (1+2\varphi )}
        \Bigg[ {1\over 2} \left( \chi^i_{\; \; |j} +\chi_j^{\; \; |i} \right)
              -{1\over 3}\delta^i_j\chi^k_{\; \; |k}
              \nonumber \\
    \fl           -{1\over 1+2\varphi }
              \left( \chi^i \varphi_{,j} + \chi_j \varphi^{|i}
                    -{2\over 3}\delta^i_j \chi^k \varphi_{,k} \right) \Bigg] \Bigg\}
       +{c\chi^k\over a^2 {\cal N}(1+2\varphi )}
       \nonumber \\
    \fl  \times \Bigg\{ {c\over a^2 {\cal N} (1+2\varphi )}
       \Bigg[ {1\over 2} \left( \chi^i_{\; \; |j} +\chi_j^{\; \; |i} \right)
              -{1\over 3}\delta^i_j\chi^l_{\; \; |l}
              -{1\over 1+2\varphi }
              \Bigg( \chi^i \varphi_{,j} + \chi_j \varphi^{|i}
                -{2\over 3}\delta^i_j \chi^l \varphi_{,l} \Bigg) \Bigg] \Bigg\}_{|k}
                \nonumber \\
    \fl     -{c^2\over a^2 (1+2\varphi )}
      \left[ {1\over 1+2\varphi } \left( \varphi^{|i}_{\; \; \; j}
       -{1\over 3}\delta^i_j\varphi^{|k}_{\; \; k}\right)
            +{1\over {\cal N}}\left( {\cal N}^{|i}_{\; \; \; j}-{1\over 3}\delta^i_j{\cal N}^{|k}_{\; \; \; k} \right) \right]
            \nonumber \\
    \fl         ={8\pi G\over c^2} (\widetilde \mu + \widetilde p )
                               \left[ {\widehat\gamma^2 \widehat v^i \widehat v_j\over c^2 (1+2\varphi )}
                                -{1\over 3}\delta^i_j  \left(\widehat\gamma^2 -1 \right) \right]
                       +{c^2 \over a^4 {\cal N}^2 (1+2\varphi )^2 }
                                \Bigg[ {1\over 2}
                                                              \Bigg( \chi^i_{\; \; |k}\chi_j^{\; \; |k}
                           -\chi_{k|j}\chi^{k|i} \Bigg)
                           \nonumber \\
     \fl  \qquad
                   +{1\over 1+2\varphi } \Bigg(\chi^{k|i}\chi_k\varphi_{,j}
                                         -\chi^{i|k}\chi_j\varphi_{,k}
                                         +\chi_{k|j} \chi^k\varphi^{|i}
                                       -\chi_{j|k}\chi^i\varphi^{|k} \Bigg)
                                       \nonumber \\
     \fl       \qquad             +{2\over (1+2\varphi )^2}
                   \left(\chi^i\chi_j \varphi^{|k}\varphi_{,k}
                    -\chi^k\chi_k \varphi^{|i}\varphi_{,j}\right)
             \Bigg]
          -{c^2\over a^2 (1+2\varphi )^2}
             \Bigg[ {3\over 1+2\varphi }
             \nonumber \\
      \fl   \qquad \times      \left( \varphi^{|i}\varphi_{,j}
                                           -{1\over 3}\delta^i_j \varphi^{|k}\varphi_{,k} \right)
                                 +{1\over {\cal N}} \left( \varphi^{|i}{\cal N}_{,j}
                                          +\varphi_{,j} {\cal N}^{|i}
                                          -{2\over 3}\delta^i_j \varphi^{|k}{\cal N}_{,k}\right)                                  \Bigg].
                                      \label{eq5}
\eea

\noindent
ADM energy conservation:
\bea
   \fl {1 \over {\cal N}} \left[ \widetilde\mu + (\widetilde\mu + \widetilde p )
                                  (\widehat\gamma^2 -1 ) \right]^\cdot
               + {c\over a^2 {\cal N}} {\chi^i\over 1+2\varphi} \left[ \widetilde\mu + (\widetilde\mu +\widetilde p )
         (\widehat\gamma^2 -1) \right]_{|i}
         \nonumber \\
   \fl             +(\widetilde \mu +\widetilde p) (3H-\kappa ){1\over 3} (4\widehat\gamma^2 -1 )
                +\Big ( {\widetilde\mu +\widetilde p \over a(1+2\varphi )}\widehat\gamma^2 \widehat v^i \Big)_{|i}
                               +\Big ( {3\varphi_{,i}\over 1+2\varphi } +2{{\cal N}_{,i}\over {\cal N}}\Big )
                 {\widetilde\mu +\widetilde p \over a(1+2\varphi )}\widehat\gamma^2 \widehat v^i
                 \nonumber \\
    \fl      \qquad       =-{\widehat\gamma^2 (\widetilde\mu +\widetilde p)\over ca^2 {\cal N} (1+2\varphi )^2}
                \Big[ \chi^{i|j}\widehat v_i \widehat v_j
                     -{1\over 3} \chi^i_{\; \; |j}\widehat v^j \widehat v_i
                                -{2\over 1+2\varphi } \Big( \widehat v^i \widehat v^j \chi_i \varphi_{,j}
                              -{1\over 3}\widehat v^i \widehat v_i\chi^j\varphi_{,j} \Big) \Big].
                          \label{eq6}
\eea

\noindent
ADM momentum conservation:
\bea
   \fl
         \left(  {1 \over {\cal N}} {\partial\over \partial t}+3H -\kappa \right)
                                  \left[ a (\widetilde\mu + \widetilde p )  \widehat\gamma^2
                                   \widehat v_i \right]
                +{c\over a^2 \cal N } {\chi^j\over (1+2\varphi )}
                  \left[ a (\widetilde\mu + \widetilde p ) \widehat\gamma^2 \widehat v_i \right]_{|j}
                  +c^2 \widetilde p_{,i}
                         +c^2 (\widetilde \mu + \widetilde p ){{\cal N}_{,i}\over {\cal N}}
                         \nonumber \\
          \fl \qquad        =- \left[ (\widetilde\mu + \widetilde p ) {\gamma^2 \widehat v^j \widehat v_i\over 1+2\varphi }\right]_{|j}
                    -{c\over a{\cal N}} \left( {\chi^j\over 1+ 2\varphi }\right)_{|i}
           (\widetilde\mu + \widetilde p) \widehat\gamma^2 \widehat v_j
           \nonumber \\
     \fl  \qquad    -{\widetilde\mu + \widetilde p \over 1+2\varphi } \widehat\gamma^2 \widehat v^j
           \left[ {1\over 1+2\varphi } (3\widehat v_i\varphi_{,j}-\widehat v_j \varphi_{,i} )
                  +{1\over {\cal N}}(\widehat v_i {\cal N}_{,j} +\widehat v_j {\cal N}_{,i} )\right] .
                                                                                \label{eq-momentum-conserve}
\eea

%Here, we used the fluid velocity $\widehat v_i$, which is defined as
%$v_i \equiv \widehat\gamma \widehat v_i$.
%$\widehat\gamma$ is the Lorentz factor:
%$\widehat \gamma \equiv  {1\over \sqrt{ 1-{\widehat v^k \widehat v_k \over c^2 (1+2\varphi )}}}$.
%We also recovered the speed of light $c$.
Instead of  the ADM-energy and momentum conservation equations, we can use the covariant energy and the momentum conservation equations; see the Appendix C in \cite{Hwang-2013}.

\noindent
Covariant energy conservation:
\bea
   \fl
   \left[ {\partial\over \partial t} + {1\over a(1+2\varphi )}
          \left( {\cal N} \widehat v^i + {c\over a}\chi^i \right) \nabla_i \right] \widetilde\mu
           +(\widetilde \mu +\widetilde p ) \Bigg\{ (3H-\kappa ){\cal N}
                       +{({\cal N}\widehat v^i )_{|i} \over a(1+2\varphi )}
                       \nonumber \\
    \fl \qquad             +{{\cal N} \widehat v^i \varphi_{,i} \over a(1+2\varphi )^2}
                       +{1\over \widehat\gamma }
                       \left[ {\partial\over \partial t} +{1\over a(1+2\varphi )}
                            \left( {\cal N} \widehat v^i +{c\over a}\chi^i \right)
                              \nabla_i  \right] \widehat\gamma
                                                     \Bigg\} =0.
                                                      \label{eq7}
\eea

\noindent
Covariant momentum conservation:
\bea
   \fl   {\partial\over \partial t} (a\widehat\gamma\widehat v_i ) + {1\over a(1+2\varphi )}
          \left( {\cal N} \widehat v^k + {c\over a}\chi^k \right) ( a\widehat\gamma\widehat v_i )_{|k}
          \nonumber \\
    \fl \qquad
          +{1\over \widetilde \mu + \widetilde p }
          \Bigg\{ c^2 {{\cal N}\over \widehat\gamma } \widetilde p_{,i}
               +a\widehat\gamma\widehat v_i \left[ {\partial\over \partial t}
           + {1\over a(1+2\varphi )}
           \left( {\cal N} \widehat v^k
                  +{c\over a} \chi^k \right) \nabla_k \right]
                  \tilde p \Bigg\}
                  \nonumber \\
    \fl  \qquad + c^2 \widehat\gamma {\cal N}_{,i}
         +{1-\widehat\gamma^2 \over \widehat\gamma} {c^2 {\cal N} \varphi_{,i}\over 1+2\varphi }
         +{c\over a} \widehat\gamma \widehat v^k \left( {\chi_k \over 1+2\varphi } \right)_{|i} = 0.
          \label{eq8}
   \eea
The equation (\ref{eq1}) is derived from the trace of the extrinsic curvature as $K \equiv - 3 H + \kappa$; ${\cal N}$ is related to the lapse function in equation (\ref{ADM-metric}), and $\overline{K}^i_j$ is the tracefree part of extrinsic curvature in equation (\ref{extrinsic-curvature}).  ${\cal N}$ and $\overline{K}^i_j \overline{K}^j_i$ are given as
\bea
   \fl {\cal N} = \sqrt{ 1 + 2 \alpha + {\chi^{k} \chi_{k} \over
       a^2 (1 + 2 \varphi)}},
       \nonumber \\
   \fl            \overline{K}^i_j \overline{K}^j_i
       = {1 \over a^4 {\cal N}^2 (1 + 2 \varphi)^2}
       \Bigg\{
       {1 \over 2} \chi^{i|j} \Bigg( \chi_{i|j} + \chi_{j|i} \Bigg)
          - {1 \over 3} \chi^i_{\;\;|i} \chi^j_{\;\;|j}
          \nonumber \\
    \fl \qquad \quad
            - {4 \over 1 + 2 \varphi} \Bigg[
       {1 \over 2} \chi^i \varphi^{|j}
            \Bigg(
       \chi_{i|j} + \chi_{j|i} \Bigg)
       - {1 \over 3} \chi^i_{\;\;|i} \chi^j \varphi_{,j} \Bigg]
       \nonumber \\
    \fl \qquad \qquad      + {2 \over (1 + 2 \varphi)^2}
        \Bigg(
       \chi^{i} \chi_{i} \varphi^{|j} \varphi_{,j}
       + {1 \over 3} \chi^i \chi^j \varphi_{,i} \varphi_{,j} \Bigg) \Bigg\}.
   \label{K-bar-eq}
\eea

Now, we have the complete set of the fully nonlinear perturbations in (\ref{eq1})-(\ref{eq8}) in a non-flat background universe.

The dimensions are,
\bea
   & & [a] = [\widetilde g_{ab}] = [\widetilde u_a]
       = [\alpha] = [\varphi] = [\chi^i]
       = [v^i/c]
       = [\widehat v^i/c]
%       = [\overline v^i/c]
       = 1,
       \nonumber \\
 & &       [v/c] = L, \quad
          [x^i] = [ c dt] \equiv [ a d \eta] = L, \quad
       [\chi] = L, \quad
       [\kappa] = T^{-1},
       \nonumber \\
   & &    [\widetilde T_{ab}] = [\widetilde \mu]
       = [\widetilde \varrho c^2]
       = [\widetilde p], \quad
       [G \widetilde \varrho] = T^{-2}, \quad
       [\overline K]=L^{-2}, \quad [\Lambda] = L^{-2}.
\eea
%%%%%%%%%%%%%%%%%%%%%%%%%%%%%%%%%%%%%%%%%%%%%%%%%%%%%%%%%%%%%%%
%
%%%%%%%%%%%%%%%%%%%%%%%%%%%%%%%%%%%%%%%%%%%%%%%%%%%%%%%%%%%%%%%

%%%%%%%%%%%%%%%%%%%%%%%%%%%%%%%%%%%%%%%%%%%%%%%%%%%%%%%%%%%%%%%
\section{Analysis in the Comoving gauge}
                                             \label{sec:CG}

We see that the background curvature term explicitly appears in the ADM energy constraint equation only.
The effect of background curvature appears in the ADM momentum constraint equation as well: see equations (\ref{eq2}) and
(\ref{eq3}).

In order to see the effects of the background curvature in detail, in this section we consider the scalar-type perturbations, so $\widehat v_i = -\widehat  v_{,i}$ and $\chi_i = \chi_{,i}$.
We take the comoving gauge
\bea
   & & \widehat v \equiv 0,
\eea
thus $\widehat v_i = 0$.
From equation (\ref{eq-momentum-conserve}) or (\ref{eq8}) we have
\bea
   & & \widetilde p_{,i}
       = - \left( \widetilde \mu + \widetilde p \right) {{\cal N}_{,i} \over {\cal N}}.
           \label{eq7-CG}
\eea
Equations (\ref{eq1})-(\ref{eq8}) are the complete set of equations for the variables $\delta$, $\kappa$, $\varphi$, $\chi$ and $\alpha$.
% As explained in Section \ref{sec:convention} and below equation (\ref{temporal-gauges-NL})
Since the comoving gauge completely fixes the gauge modes, all perturbation variables in the comoving gauge are gauge invariant even to the nonlinear order: see Section 2 in \cite{Hwang-2013}.
%\bea
%   & & \delta = \delta_v, \quad
%      \kappa = \kappa_v, \quad
%      \varphi = \varphi_v, \quad
%      \chi = \chi_v, \quad
%      \alpha = \alpha_v.
%\eea

Subtracting the background equation, equations (\ref{eq6}) or (\ref{eq7}) gives
\bea
   & & \kappa = {1 \over \widetilde \mu + \widetilde p} {1 \over {\cal N}}
       \left( {\partial \over \partial t}
       + {c\chi^{|i} \over a^2 ( 1 + 2 \varphi )} \nabla_i \right)
       {\widetilde \mu}
       - {\dot \mu \over \mu + p}. \label{kappa}
\eea
Equation (\ref{eq4}) gives
\bea
  \fl  {1 \over {\cal N}}
     \dot\kappa + 2H\kappa
         - 3 \left( {1 \over {\cal N}} - 1 \right) \dot H
       - {4 \pi G\over c^2} \left( \delta \mu + 3 \delta p \right)
       + {c^2{\cal N}^{|i}_{\; \; \; i} \over a^2 {\cal N} (1 + 2 \varphi)}
       \nonumber \\
   \fl \qquad
       = {1 \over 3} \kappa^2
       - {c \over a^2 {\cal N} (1 + 2 \varphi)} \left(
       \chi^{|i} \kappa_{,i}
             +c {\varphi^{|i} {\cal N}_{,i} \over 1 + 2 \varphi} \right)
              + c^2\overline{K}^i_j \overline{K}^j_i.    \label{diff-eq}
\eea
Combining these we have a second-order differential equation for $\widetilde \mu$.

%%%%%%%%%%%%%%%%%%%%%%%%%%%%%%%%%%%%%%%%%%%%%%%%%%%%%%%%%%%%%%%
\subsection{Zero-pressure case}

Now, we assume the pressure is negligible, then ${\cal N}_{,i} = 0$ from equation (\ref{eq7-CG}). Thus, we have ${\cal N} = 1$ and
\bea
   & & \alpha
       = - {1 \over 2} {\chi^{|k} \chi_{,k} \over a^2 (1 + 2 \varphi)}.
   \label{alpha-CG} \\
   \nonumber
\eea
Notice that even in the zero-pressure case, with our choice of the spatial gauge condition  $\gamma = 0$,
to the nonlinear order the comoving gauge ($\widehat v = 0$) does not imply the
synchronous gauge ($\alpha = 0$) \cite{Hwang-2006}.

Using equations (\ref{eq4}), (\ref{eq6}), (\ref{eq2}) and (\ref{eq3}) we have the complete set of equations for $\delta$ and $\kappa$.
Equation (\ref{kappa}) and (\ref{diff-eq}), respectively, give
\bea
   & &         \dot \delta - \kappa = \delta \kappa
       - {c\chi^{|i} \delta_{,i} \over a^2 (1 + 2 \varphi)},
   \label{eq6-CG-mde} \\
   & &        \dot \kappa
       + 2 H \kappa
       - {4 \pi G\over c^2}  \delta \mu
         = {1 \over 3} \kappa^2
       - {c\chi^{|i} \kappa_{,i} \over a^2 (1 + 2 \varphi)}
       + c^2\overline{K}^i_j \overline{K}^j_i.
   \label{eq4-CG-mde}
\eea
From these we have
\bea
   \fl \ddot \delta
       + 2 H \dot \delta
       - {4 \pi G \mu\over c^2} \delta
          = {1 \over a^2} \left( a^2 \kappa \delta \right)^{\displaystyle\cdot}
       - {c \over a^2} \left( {\chi^{|i} \delta_{,i} \over 1 + 2 \varphi} \right)^{\displaystyle\cdot}
       + {1 \over 3} \kappa^2
       - {c\chi^{|i} \kappa_{,i} \over a^2 (1 + 2 \varphi)}
       + c^2\overline{K}^i_j \overline{K}^j_i.
\eea
The variables $\chi$ and $\varphi$ are obtained from equation (\ref{eq3}) and (\ref{eq2}), respectively
\bea
   \fl \kappa_{,i}
       + { c\over a^2 (1+2\varphi )}\Bigg[ \left( \Delta
        + 3{\overline K}\right) \chi \Bigg]_{,i}
        \nonumber \\
    \fl \qquad   =
       {c \over a^2 ( 1 + 2 \varphi)^2}
             \Bigg[
                           {3\over 2} \left( \chi^{|j}\varphi_{,i|j}
         +\chi_{,i|j}\varphi^{|j}
         -{2\over 3} \chi^{|j}_{\; \; \; i}\varphi_{,j}
         -{2\over 3} \chi^{|j}\varphi_{,j|i} \right)
         \nonumber \\
     \fl  \qquad  \quad    +2  \chi^{|j}_{\; \; \; j}  \varphi_{,i}
                                 -{3\over 2} \varphi_{,j}  \chi^{|j}_{\; \; \; i}
    + {3 \over 2} \chi_{,i}  \varphi^{|j}_{\; \; \; j}
       - {3 \over 2} {1 \over 1 + 2 \varphi}
       \left( \chi_{,i} \varphi_{,j}
       + {1 \over 3} \chi_{,j} \varphi_{,i} \right) \varphi^{|j} \Bigg],
                 \label{eq3-CG-mde} \\
    \fl - {3 \over 2} \left( H^2
       - {8 \pi G \over 3c^2} \mu +{c^2\overline K\over a^2 (1+2\varphi )} - {\Lambda c^2\over 3} \right)
       +
       H \kappa
       + {4 \pi G\over c^2} \mu \delta
       + { c^2\varphi^{|i}_{\; \; \; i} \over a^2 (1 + 2 \varphi)^2}
       \nonumber \\
    \fl \qquad   = {1 \over 6} \kappa^2
       + {3 \over 2} {c^2\varphi^{|i} \varphi_{,i} \over a^2 (1 + 2 \varphi)^3}
       - {c^2 \over 4} \overline{K}^i_j \overline{K}^j_i
       .
   \label{eq2-CG-mde}
\eea
To the linear order, using equations (\ref{eq1}) and (\ref{eq3-CG-mde}) we have
\bea
   & & \dot\varphi = {{\overline K} c\over a^2} \chi .
\eea
Thus, in the presence of the background curvature, $\varphi$ in the comoving gauge is not conserved
even to the linear order, while in a flat background, we have $\dot\varphi =0$.

%We have the equation for $\dot \varphi$  from equations
%(\ref{eq1}) and (\ref{eq3-CG-mde}) as
%\bea
%   & & \left[ \ln{ \left( 1 + 2 \varphi \right) } \right]^{\displaystyle\cdot}_{,i}
%       = - {c \over a^2 ( 1 + 2 \varphi )^2}
%       \left[ \chi^{,k} \varphi_{,i|k}
%       + \chi_{,i}  \varphi^{,j}_{\; \; |j}
%       - {1 \over 1 + 2 \varphi}
%       \left( \chi_{,i} \varphi_{,k}
%       + 3 \chi_{,k} \varphi_{,i} \right) \varphi^{,k} \right]
%                                             .
%\eea
%thus $\dot \varphi = 0$ to the linear order.

%%%%%%%%%%%%%%%%%%%%%%%%%%%%%%%%%%%%%%%%%%%%%%%%%%%%%%%%%%%%%%%
%\subsection{Newtonian correspondence}
\subsection{Pure relativistic corrections to fully nonlinear order}

In order to see the pure Einstein's gravity corrections, we arrange Equations (\ref{eq6-CG-mde}), (\ref{eq4-CG-mde}) and (\ref{eq3-CG-mde}) as
\bea
   \fl
       \dot \delta - \kappa
       - \delta \kappa
       + {c \over a^2} \chi^{|i} \delta_{,i}
       = {2c \varphi \chi^{|i} \delta_{,i} \over a^2 (1 + 2 \varphi)},
   \label{eq6-CG-mde-pure-GR} \\
   \fl
       \dot \kappa
       + 2 H \kappa
       - {4 \pi G \over c^2}\delta \mu
       - {1 \over 3} \kappa^2
       + {c \over a^2} \chi^{|i} \kappa_{,i}
       - {c^2 \over a^4} \left[
       \chi^{|ij} \chi_{,i|j}
       - {1 \over 3} \left( \Delta \chi \right)^2 \right]
       \nonumber \\
   \fl \qquad =
       {2c \varphi \chi^{|i} \kappa_{,i} \over a^2 (1 + 2 \varphi)}
       - {4c^2 \varphi ( 1 + \varphi ) \over a^4 (1 + 2 \varphi)^2}
       \left[ \chi^{|ij} \chi_{,i|j}
       - {1 \over 3} \left( \Delta \chi \right)^2 \right]
   \nonumber \\
   \fl \qquad       + {2c^2 \over a^4 (1 + 2 \varphi)^3}
       \Bigg\{
       {2 \over 3} \left( \Delta \chi \right) \chi^{|i} \varphi_{,i}
       - 2 \chi^{|ij} \chi_{,i} \varphi_{,j}
       + {1 \over 1 + 2 \varphi} \Bigg[
       {1 \over 3} \Bigg( \chi^{|i} \varphi_{,i} \Bigg)^2
       + \chi^{|i} \chi_{,i} \varphi^{|j} \varphi_{,j} \Bigg] \Bigg\},
   \label{eq4-CG-mde-pure-GR} \\
   \fl \kappa_{,i}+
   { c
   \over a^2 }\Bigg[\Bigg( \Delta+3{\overline K}\Bigg)\chi\Bigg]_{,i}
           ={2c\varphi\over a^2 (1+2\varphi)}
              \Bigg[\Bigg( \Delta+3{\overline K}\Bigg)\chi\Bigg]_{,i}
              \nonumber \\
        \fl \qquad
                     + {c \over a^2 ( 1 + 2 \varphi)^2}
                    \Bigg[
                           {3\over 2} \left( \chi^{|j}\varphi_{,i|j}
         +\chi_{,i|j}\varphi^{|j}
         -{2\over 3} \chi^{|j}_{\; \; \; i}\varphi_{,j}
         -{2\over 3} \chi^{|j}\varphi_{,j|i} \right)
                  +2   \chi^{|j}_{\; \; \; j}  \varphi_{,i}
                   \nonumber \\
         \fl           \qquad -{3\over 2} \varphi_{,j} \chi^{|j}_{\; \; \; i}
    + {3 \over 2} \chi_{,i}  \varphi^{|j}_{\; \; \; j}
       - {3 \over 2} {1 \over 1 + 2 \varphi}
       \left( \chi_{,i} \varphi_{,j}
       + {1 \over 3} \chi_{,j} \varphi_{,i} \right) \varphi^{|j} \Bigg].
          \label{eq3-CG-mde-pure-GR}
\eea
The pure relativistic corrections appear through $\varphi$, the spatial curvature perturbation in the comoving gauge,
and these are presented in the right-hand-side of the equations.
The background curvature can be regarded as the pure relativistic correction as well.

%%%%%%%%%%%%%%%%%%%%%%%%%%%%%%%%%%%%%%%%%%%%%%%%%%%%%%%%%%%%%%%
\subsection{Relativistic/Newtonian correspondence}
                                             \label{sec:correspondence}

In the flat background, we found that Equations (\ref{eq6-CG-mde-pure-GR})-(\ref{eq4-CG-mde-pure-GR}) without $\varphi$ exactly coincide with the Newtonian hydrodynamic equations of the mass and the momentum conservation as shown in Section 5.3 of \cite{Hwang-2013}. %(removing the gravitational potential in the momentum conservation equation using the Poisson's equation), respectively.
% This statement is true to fully nonlinear order in perturbation in the presence of the cosmological %constant in the background.
In this section, in order to see the relativistic/Newtonian correspondence in the presence of the background curvature, we consider the weak gravity limit, thus neglect
$\varphi$ terms. Equations (\ref{eq6-CG-mde-pure-GR})-(\ref{eq3-CG-mde-pure-GR}) give
% and (\ref{eq3-CG-mde}),
\bea
   & & \!\!\!\!\!
       \dot \delta - \kappa = \delta \kappa
       - {c \over a^2} \chi^{,i} \delta_{,i},
   \label{eq6-CG-mde-2} \\
   & & \!\!\!\!\!
       \dot \kappa
       + 2 H \kappa
       - {4 \pi G\over c^2}  \delta \mu
          = {1 \over 3} \kappa^2
       - {c \over a^2} \chi^{,i} \kappa_{,i}
       + {c^2 \over a^4} \left[
       \chi^{,i|j} \chi_{,i|j}
       - {1 \over 3} \left( \Delta \chi \right)^2 \right],
   \label{eq4-CG-mde-2} \\
   & & \!\!\!\!\!
       \kappa + c{\Delta +3\overline K \over a^2} \chi = 0.
   \label{eq3-CG-mde-2}
\eea
We identify $\delta$ and ${\bf u}$ as the Newtonian density and velocity perturbations with
\bea
   \kappa \equiv - {1 \over a} \nabla \cdot {\bf u}\equiv -{\Delta\over a}u,
\eea
where ${\bf u} \equiv \nabla u$.
Equation (\ref{eq3-CG-mde-2}) can be arranged to give
\bea
    & & {\Delta \over a} (u-c{\chi\over a} ) = 3c{{\overline K}\over a^2}\chi .
\eea
We can rewrite the equations (\ref {eq6-CG-mde-2}), (\ref {eq4-CG-mde-2}) as
 \bea
 \fl \dot \delta + {1 \over a} \nabla \cdot {\bf u}
       + {1 \over a} \nabla \cdot \left( \delta {\bf u} \right)
       =
       {1\over a}(\nabla\delta )\cdot \nabla (u-{c\chi\over a} ),
        \label{eq4-CG-mde-Newton1}
\eea
\bea
  \fl    {1 \over a} \nabla \cdot \left( \dot {\bf u} + H {\bf u} \right)
       + 4 \pi G \varrho \delta
      +{1 \over a^2} \nabla \cdot \left( {\bf u} \cdot \nabla {\bf u} \right)
            \nonumber \\
  \fl    \qquad =
             {1\over a^2} \left(u-{c\chi\over a}\right)^{|ij}\left(u+{c\chi\over a}\right)_{|ij}
             +{2{\overline K}\over a^2} u^{|i}u_{,i}
      \nonumber \\
  \fl \qquad
 -{1\over 3}{1\over a^2}
       \left[ \Delta \left(u-{c\chi\over a} \right)\right]\Delta\left( u+{c\chi\over a}\right)
        +{1\over a^2}\left( u-{c\chi\over a}\right)^{|i}
 \left[ \Delta \left(u+{c\chi\over a}\right)\right]_{|i},
 \label{eq4-CG-mde-Newton2}
\eea
where we have located the background curvature contribution in right-hand-side of the equations.
In the presence of the background curvature, we  have additional relativistic
contribution from the background curvature to the Newtonian equations.
Notice that the right-hand-sides are the second order.

In the case of the flat background, the relativistic and Newtonian correspondence was presented in \cite{Hwang-2013}.
Up to the second order, the perturbation equations including the background curvature have been
investigated in our previous work \cite{NL-2007}.
The background curvature effect also appears from the second order in perturbation.
%We note that the dimension of $\chi$ was defined differently in \cite{NL-2007}.

The pure Einstein's gravity contributions appear in terms of the background curvature and $\varphi$ to the fully nonlinear order.  We have shown that the $\varphi$ terms start to appear from the third order perturbation, thus in a flat background  we have exact relativistic/Newtonian correspondences of the density and velocity perturbations to the second-order perturbation \cite{NL-2004}.

%
%%%%%%%%%%%%%%%%%%%%%%%%%%%%%%%%%%%%%%%%%%%%%%%%%%%%%%%%%%%%%%%
%%%%%%%%%%%%%%%%%%%%%%%%%%%%%%%%%%%%%%%%%%%%%%%%%%%%%%%%%%%%%%%
\section{Minimally coupled scalar field}
                                               \label{sec:MSF}

In this section we present the fully nonlinear and exact formulation for a
minimally coupled scalar field.

We consider a minimally coupled scalar field $\widetilde \phi$. The energy and momentum tensor is
\bea
   & & \widetilde T_{ab} = \widetilde \phi_{,a} \widetilde \phi_{,b}
       - \left[ {1 \over 2} \widetilde \phi^{;c} \widetilde \phi_{,c} + \widetilde V (\widetilde \phi) \right] \widetilde g_{ab},
   \label{Tab-MSF}
\eea and the equation of motion is
\bea
   & & \widetilde \phi^{;c}_{\;\;\; c} = \widetilde V_{,\widetilde \phi}.
   \label{EOM-MSF}
\eea 
Equation (\ref{fluids}) gives the fluid quantities. In this section, we set $c\equiv 1$.

Since the minimally coupled scalar field has no anisotropic stress, the same fluid formulation remains valid:  the fluid quantities are replaced by the ones in terms of the scalar field.

From Equations (\ref{Tab}) and (\ref{Tab-MSF}) we have
\bea
   & & \widetilde u_i
       = - {1 \over \widetilde \mu} \widetilde T_{ib} \widetilde u^b
       = - {\widetilde \phi_{,i} \over \widetilde \phi_{,c} \widetilde u^c}, \quad
       \widetilde h^b_a \widetilde \phi_{,b} = 0, \quad
       0 = \widetilde h^{cd} \widetilde \phi_{,c} \widetilde \phi_{,d}
       = \widetilde \phi^{;c} \widetilde \phi_{,c}
       + {\widetilde {\dot {\widetilde \phi}}}^2,
\eea
where
$\widetilde h_{ab} \equiv \widetilde g_{ab}+\widetilde u_a \widetilde u_b $ is the
projection tensor and
$ {\widetilde {\dot {\widetilde \phi}}} \equiv \widetilde \phi_{,c} \widetilde u^c $.
We set $\widetilde \phi = \phi + \delta \phi$, where $\phi (\eta)$ is
the background order scalar field, and the $\delta\phi$ is the perturbed part with arbitrary amplitude .

We obtain
\bea
   & & {\widetilde {\dot {\widetilde \phi}}}
       = \widetilde \phi_{,c} \widetilde u^c
       = {\widetilde \phi_{,i} \widehat\gamma\widehat v^i \over a ( 1 + 2 \varphi )}
       + \widehat \gamma {D \widetilde \phi \over D t}
       = {1 \over \widehat \gamma} {D \widetilde \phi \over D t},
%       , \quad
%       a \widehat\gamma^2 \widehat v^k \widehat v_k {D \widetilde \phi \over D t}
%       = - \widehat \gamma^2 \widehat v^i \widetilde \phi_{,i},
%   \nonumber \\
%   & & \widetilde \phi^{;c} \widetilde \phi_{,c}
%       = \widetilde g^{cd} \widetilde \phi_{,c} \widetilde \phi_{d}
%            = - {1 \over {\cal N}^2} \dot {\widetilde \phi}^2
%       - 2 {\chi^i \over a^2 {\cal N}^2 (1 + 2 \varphi)}
%       \dot {\widetilde \phi} \widetilde \phi_{,i}
%       \nonumber \\
%    & &  \qquad  + {1 \over a^2 (1 + 2 \varphi)}
%       \left( g^{(3)ij}
%       - {\chi^i \chi^j \over a^2 {\cal N}^2 (1 + 2 \varphi)} \right)
%       \widetilde \phi_{,i} \widetilde \phi_{,j},
\eea where we define
\bea
   & & {D \over Dt}\widetilde\phi
       \equiv {1 \over {\cal N}}
       \left( {\partial \over \partial t}
       + {\chi^i \over a^2 (1 + 2 \varphi)} \nabla_i \right)\widetilde\phi .
%       = {1 \over N} \left( \partial_0
%       - N^i \nabla_i \right)\widetilde\phi.
\eea
From equations (\ref{Tab-MSF}) and (\ref{fluids}) we have
\bea
   & & \widetilde \mu = {1 \over 2} {\widetilde {\dot {\widetilde \phi}}}^2
       + \widetilde V, \quad
       \widetilde p = {1 \over 2} {\widetilde {\dot {\widetilde \phi}}}^2
       - \widetilde V, \quad
       a\widehat\gamma \widehat v_i {\widetilde {\dot {\widetilde \phi}}} = - \widetilde \phi_{,i}.
   \label{fluids-MSF}
\eea
Equations (\ref{velocity}) and (\ref{fluids-MSF}) provide
\bea
  \fl \widehat v = {1 \over a} \Delta^{-1}
       \left( \widetilde \phi^{|k} / ({D\widetilde\phi\over Dt})
       \right)_{|k}, \quad
      \widehat v_i^{(v)}
       = - {1 \over a} \widetilde \phi_{,i} /  ({D\widetilde\phi\over Dt})
       + {1 \over a} \left[ \Delta^{-1}
       \left( \widetilde \phi^{|k} /  ({D\widetilde\phi\over Dt})
       \right)_{,k}\right]_{,i}.
   \label{v}
\eea
We also have
\bea
 & &  \widehat\gamma = {1\over \sqrt{1-{\widetilde\phi^{|k}\widetilde\phi_{,k}\over a^2 (1+2\varphi ) (D\widetilde\phi/Dt)^2}}}.
\eea
From equation  (\ref{EOM-MSF}) we have \bea
   & & - \widetilde \phi^{;c}_{\;\;\; c}
       = {D^2 \widetilde \phi \over Dt^2}
       + \left( 3 H
       - \kappa \right)
       {D \widetilde \phi \over Dt}
       - {( {\cal N} \sqrt{1 + 2 \varphi}
       \widetilde \phi^{|i} )_{|i} \over a^2 {\cal N} (1 + 2 \varphi)^{3/2}}
       = - \widetilde V_{,\widetilde \phi}
       ( \widetilde \phi ).
   \label{EOM-perturbed}
\eea
Equation (\ref{EOM-perturbed}) gives
\bea
  \fl \ddot {\widetilde \phi}
       + \left( 3 H {\cal N}
       - {\cal N} \kappa
       - {\dot {\cal N} \over {\cal N}}
       - {\chi^i {\cal N}_{|i} \over
       a^2 {\cal N} ( 1 + 2 \varphi )} \right) \dot {\widetilde \phi}
       + {2 \chi^i \over
       a^2 ( 1 + 2 \varphi )} \dot {\widetilde \phi}_{,i}
       \nonumber \\
   \fl   \qquad  - {1 \over a^2 (1 + 2 \varphi)}
       \left( {\cal N}^2 g^{(3)ij}
       - {\chi^i \chi^j \over a^2 (1 + 2 \varphi)} \right) \widetilde \phi_{,i|j}
             + \Bigg[ - {{\cal N}^2 \over a^2 (1 + 2 \varphi)}
              \left( {{\cal N}^{|i} \over {\cal N}}
       + {\varphi^{|i} \over 1 + 2 \varphi} \right)
       \nonumber \\
   \fl \qquad    + \left( 3 H {\cal N}
       - {\cal N} \kappa
       - {\dot {\cal N} \over {\cal N}}
       - {\chi^k {\cal N}_{|k} \over
       a^2 {\cal N} ( 1 + 2 \varphi )} \right)
       {\chi^i \over a^2 (1 + 2 \varphi)}
       \nonumber \\
   \fl  \qquad     + \left( {\chi^i \over a^2 (1 + 2 \varphi)}
       \right)^{\displaystyle\cdot}
          + {\chi^k \over a^4 (1 + 2 \varphi)}
       \left( {\chi^i \over 1 + 2 \varphi} \right)_{|k}
       \Bigg] \widetilde \phi_{,i} \nonumber \\
    \fl \qquad       = - {\cal N}^2 \widetilde V_{,\widetilde \phi}
       ( \widetilde \phi ).
   \label{EOM}
\eea
To the background order, we have \bea
   & & \mu = {1 \over 2} \dot \phi^2 + V, \quad
       p = {1 \over 2} \dot \phi^2 - V,
   \label{fluids-BG}
\eea and \bea
   & & \ddot \phi + 3 H \dot \phi + V_{,\phi} = 0.
   \label{EOM-BG-order}
\eea
The entropic perturbation $e$ is given by
 \bea
   & & e \equiv \delta p - {\dot p \over \dot \mu} \delta \mu
       = \left( 1 + {\ddot \phi \over 3 H \dot \phi} \right)
       \left( {\widetilde {\dot {\widetilde \phi}}}^2 - \dot \phi^2 \right)
       + {2 \ddot \phi \over 3 H \dot \phi}
       \left( \widetilde V - V \right).
\eea
%In the comoving gauge
%\bea
%   & & \widetilde u_i = - {1 \over \widetilde \mu} \widetilde T_{ib} \widetilde u^b
%       = - {\widetilde \phi_{,i} \over \widetilde \phi_{,c} \widetilde u^c}.
%\eea

From equation (\ref{v}) we notice that $\delta\phi =0$ implies $\widehat v_i =0$,
and $\widehat v=0$ also implies $\widetilde\phi$ = const in space, thus $\delta\phi =0$ as well.
Thus, to the fully nonlinear order, the comoving gauge ($v \equiv 0$) implies the uniform-field gauge ($\delta \phi \equiv 0$)
and {\it vice versa}.

In the comoving gauge we have
\bea
   & & \widetilde \mu = {1 \over 2 {\cal N}^2} \dot \phi^2
       + V, \quad
       \widetilde p = {1 \over 2 {\cal N}^2} \dot \phi^2
       - V.
\eea
To the background order, we obtain
\bea
   & & \mu = {1 \over 2} \dot \phi^2 + V, \quad
       p = {1 \over 2} \dot \phi^2 - V,
\eea and to the fully nonlinear order, we obtain
\bea
   & & \delta p = \delta \mu = {\dot\phi^2\over 2} \left({1 \over  {\cal N}^2}-1\right) .
\eea
Now, we see that in the comoving gauge the ideal fluid equations in equations (\ref{eq1})-(\ref{eq8}) remain valid with the perturbed equation of state given as $\delta p = \delta \mu$.

%%%%%%%%%%%%%%%%%%%%%%%%%%%%%%%%%%%%%%%%%%%%%%%%%%%%%%%%%%%%%%%
%
%  Discussion
%
\section{Discussion}
                           \label{sec:Discussion}

In this work, we have extended our recent work on the fully nonlinear cosmological perturbation \cite{Hwang-2013} to include the background curvature.
The equations are presented in a gauge-ready form, so the temporal gauge conditions are not fixed and
can be applied easily depending on the problem to solve.
The background curvature terms appear explicitly only in the energy and momentum constraint equations.
As an application we considered a zero-pressure irrotational fluid in the comoving gauge.
We show that in the presence of the
background curvature the pure general relativistic correction appears from the second order.
Also, we present the exact and fully nonlinear equations for the minimally coupled scalar field
in the non-flat background.
%Also, up to the second order, our fully non-linear perturbation equations show the exact
%coincidence with the previous work with the background spatial curvature.

%%%%%%%%%%%%%%%%%%%%%%%%%%%%%%%%%%%%%%%%%%%%%%%%%%%%%%%%%%%%%%%
%
% Acknowledgments
%
%%%%%%%%%%%%%%%%%%%%%%%%%%%%%%%%%%%%%%%%%%%%%%%%%%%%%%%%%%%%%%%
\section*{Acknowledgments}

We wish to thank J. Hwang for useful and important suggestions.
H.N.\ was supported by grant No.\ 2012 R1A1A2038497 from NRF.
% J.H.\ was supported by KRF Grant funded by the Korean Government (KRF-2008-341-C00022).

%%%%%%%%%%%%%%%%%%%%%%%%%%%%%%%%%%%%%%%%%%%%%%%%%%%%%%%%%%%%%%%%%
%
%   References
%
%%%%%%%%%%%%%%%%%%%%%%%%%%%%%%%%%%%%%%%%%%%%%%%%%%%%%%%%%%%%%%%%%

%%%%%%%%%%%%%%%%%%%%%%%%%%%%%%%%%%%%%%%%%%%%%%%%%%%%%%%%%%%%%%%
%%%%%%%%%%%%%%%%%%%%%%%%%%%%%%%%%%%%%%%%%%%%%%%%%%%%%%%%%%%%%%%
\section*{Appendix: Quantities useful for derivation of fully nonlinear perturbations}
                                             \label{sec:NL-eqs}

Here we present the useful quantities for deriving the fully nonlinear perturbation equations. The ADM and the covariant equations are presented in the Appendices A and C of \cite{Hwang-2013}.
Our metric is given by  \bea
   & & \widetilde g_{00} = - a^2 \left( 1 + 2 \alpha \right), \quad
       \widetilde g_{0i} = - a \chi_{i}, \quad
       \widetilde g_{ij} = a^2 \left( 1 + 2 \varphi \right) g^{(3)}_{ij}.
   \label{metric-A}
\eea 
The scale factor $a$ is a function of conformal time ($x^0 = \eta$), whereas $\alpha$, $\chi_i$ and $\varphi$ are general functions of space and time with
arbitrary amplitude.
The inverse metric is given by \bea
   & & \widetilde g^{00}
       = - {1 \over a^2} {1 + 2 \varphi \over (1 + 2 \varphi) (1 + 2 \alpha)
       + \chi^{k} \chi_{k} / a^2},
       \nonumber \\
   & &     \widetilde g^{0i} = - {1 \over a^2} {\chi^{i}/a \over (1 + 2 \varphi) (1 + 2 \alpha)
       + \chi^{k} \chi_{k} / a^2},
   \nonumber \\
   & &
       \widetilde g^{ij} = {1 \over a^2 ( 1 + 2 \varphi)}
       \left( g^{(3)ij} - { \chi^{i} \chi^{j} / a^2 \over
       (1 + 2 \varphi) (1 + 2 \alpha)
       + \chi^{k} \chi_{k} / a^2} \right).
\eea
The ADM metric can be obtained by
\bea
   & & N = a \sqrt{ 1 + 2 \alpha + {\chi^{k} \chi_{k} \over
       a^2 (1 + 2 \varphi)}}
       \equiv a {\cal N}, \quad
       N_i = - a \chi_{i}, \quad
       N^i = - {\chi^{i} \over a (1 + 2 \varphi)}, 
       \nonumber \\
   & &  \qquad    
          h_{ij} = a^2 \left( 1 + 2 \varphi \right) g^{(3)}_{ij}, \quad
       h^{ij} = {1 \over a^2 (1 + 2 \varphi)} g^{(3)ij},
   \label{ADM-metric}
\eea thus
\bea
   & & \widetilde g^{00}
       = - {1 \over a^2 {\cal N}^2}, \quad
       \widetilde g^{0i} = - {\chi^{i} \over a^3 {\cal N}^2 (1 + 2 \varphi)},
       \nonumber \\
    & &  \widetilde g^{ij} = {1 \over a^2 ( 1 + 2 \varphi)}
       \left( g^{(3)ij} - { \chi^{i} \chi^{j} \over a^2 {\cal N}^2
       (1 + 2 \varphi)} \right),
\eea and
\bea
   & & \widetilde n_i \equiv 0, \quad
       \widetilde n_0 = - a {\cal N}, \quad
       \widetilde n^i = {\chi^i \over a^2 {\cal N} ( 1 + 2 \varphi )}, \quad
       \widetilde n^0 = {1 \over a {\cal N}}.
\eea
The intrinsic three-space connection and curvatures are given by
\bea
   & & \Gamma^{(h)i}_{\;\;\;\;\;jk}
       = \Gamma^{(3)i}_{\;\;\;\;\;jk}+{1 \over 1 + 2 \varphi} \left(
       \varphi_{,j} \delta^i_k
       + \varphi_{,k} \delta^i_j
       - \varphi^{|i} g^{(3)}_{jk} \right),
       \nonumber \\
   & &     \Gamma^{(h)k}_{\;\;\;\;\;ik}
       = \Gamma^{(3)k}_{\;\;\;\;\;ik}+{3 \varphi_{,i} \over 1 + 2 \varphi},
       \nonumber \\
   & & R^{(h)i}_{\; \; \; \; \; \; jk\ell} =   R^{(3)i}_{\;\; \; \; \;\;   jk\ell}
       +{1\over 1+2\varphi }
               \left(
        \varphi_{,j|k} \delta^i_\ell -\varphi^{|i}_{\; \; \; k}g^{(3)}_{j\ell}
         -\varphi_{,j|\ell}\delta^i_k +\varphi^{|i}_{\; \; \; \ell}g^{(3)}_{jk}
                  \right)
                  \nonumber \\
   & &   \qquad   \quad       +{1\over (1+2\varphi )^2}
                           \Bigg[
         -3\varphi_{,k} \varphi_{,j} \delta^i_\ell + 3\varphi^{|i}\varphi_{,k}g^{(3)}_{j\ell}
                  +3\varphi_{,j}\varphi_{,\ell} \delta^i_k
         -3\varphi^{|i} \varphi_{,\ell} g^{(3)}_{jk}
         \nonumber \\
   & &    \qquad  \quad  +\varphi^{|m} \varphi_{,m}\Bigg(-\delta^i_k g^{(3)}_{j\ell}
         +  \delta^i_\ell g^{(3)}_{jk} \Bigg)\Bigg],
            \nonumber \\
   & & R^{(h)}_{ij}
       = R^{(3)}_{ij}- {\varphi_{,i|j} \over 1 + 2 \varphi}
       + 3 {\varphi_{,i} \varphi_{,j} \over (1 + 2 \varphi)^2}
       - \left( {\varphi^{|k}_{\; \; \; k} \over 1 + 2 \varphi}
       - {\varphi^{|k} \varphi_{,k} \over (1 + 2 \varphi)^2} \right) g^{(3)}_{ij},
    \nonumber \\
    & &  R^{(h)} = {2 \over a^2 (1 + 2 \varphi)^2}
       \left( - 2  \varphi^{|k}_{\; \; \; k}
       + 3 {\varphi^{|k} \varphi_{,k} \over 1 + 2 \varphi} \right)
       +{1\over a^2 (1+2\varphi )}  R^{(3)},
   \nonumber \\
   & & \overline{R}^{(h)i}_{\;\;\;\;\;j}
       = {1 \over a^2 (1 + 2 \varphi)^2} \left[
       - \varphi^{|i}_{\;\; \; j}
       + 3 {\varphi^{|i} \varphi_{,j} \over 1 + 2 \varphi}
       - {1 \over 3} \delta^i_j \left( -  \varphi^{|k}_{\; \; \; k}
       + 3 {\varphi^{|k} \varphi_{,k} \over 1 + 2 \varphi} \right) \right],
   \label{intrinsic-curvature}
\eea
with
\bea
    & & R^{(3)i}_{\;\; \; \; \; \;  jk\ell} = {1\over 6} R^{(3)} \left( \delta^i_k g^{(3)}_{j\ell}
                                                     -\delta^i_\ell g^{(3)}_{jk} \right),
                                                             \quad  R^{(3)}_{ij} = {1\over 3} R^{(3)}g^{(3)}_{ij},
                                                      \quad        R^{(3)} = 6\overline K.
   \label{background-curvature}
\eea
It is convenient to have
\bea
  & & B^i_{\; \; |jk} = B^i_{\; \; |kj} -R^{(3)i}_{\; \; \; \; \; \; \ell jk}B^\ell, \quad
      B_{i|jk} = B_{i|kj} + R^{(3)\ell}_{\; \; \; \; \; \; ijk}B_\ell .
\eea
The extrinsic curvature gives
\bea
   & & K_{ij}
       = - {a^2 \over {\cal N}} \Bigg[ \left( H + \dot \varphi + 2 H \varphi \right) g^{(3)}_{ij}
                + {1 \over 2 a^2} \left( \chi_{i|j} + \chi_{j|i} \right)
                \nonumber \\
   & &  \qquad   - {1 \over a^2 (1 + 2 \varphi)} \left(
       \chi_{i} \varphi_{,j}
       + \chi_{j} \varphi_{,i}
       - \chi^{k} \varphi_{,k} g^{(3)}_{ij} \right) \Bigg],
   \nonumber \\
   & & K
       = - {1 \over {\cal N} (1 + 2 \varphi)}
       \left[ 3 \left( H + \dot \varphi + 2 H \varphi \right)
       + {1 \over a^2} \chi^k_{\;\;|k}
       + {\chi^{k} \varphi_{,k} \over a^2 (1 + 2 \varphi)}
              \right]
              \nonumber \\
    & & \quad
       \equiv - 3 H + \kappa,
   \nonumber \\
   & & \overline{K}^i_j\equiv K^i_j -{1\over 3}\delta^i_j K
         = - {1 \over a^2 {\cal N} (1 + 2 \varphi)} \Bigg[
       {1 \over 2} \left( \chi^{i}_{\;\;|j} + \chi_j^{\; \; |i} \right)
       - {1 \over 3} \delta^i_j \chi^k_{\;\;|k}
       \nonumber \\
    & & \qquad
         - {1 \over 1 + 2 \varphi} \left(
       \chi^{i} \varphi_{,j}
       + \chi_{j} \varphi^{,i}
       - {2 \over 3} \delta^i_j \chi^{k} \varphi_{,k} \right)
       \Bigg].
   \label{extrinsic-curvature}
\eea
The fluid four-vector is \bea
   & & \widetilde u_i \equiv {a\widehat\gamma\widehat v_{i}\over c}, \quad
       \widetilde u_0 =
       - a {\cal N} \widehat\gamma
       - {\chi^{k} \widehat\gamma \widehat v_{k} \over c(1 + 2 \varphi)},
   \nonumber \\
   & & \widetilde u^i = {\widehat\gamma \widehat v^{i} \over c a (1 + 2 \varphi)}
       + {\widehat\gamma\chi^{i} \over a^2 {\cal N} (1 + 2 \varphi)}
       , \quad
       \widetilde u^0
       = {1 \over a {\cal N}} \widehat\gamma .
   \label{four-vector}
\eea  For $\widehat v_i = 0$ we have $\widetilde u_a = \widetilde n_a$.
The energy-momentum tensor of an ideal fluid is given by
\bea
   & & \widetilde T^0_0
       = - \widetilde \mu
       - {\widetilde \mu + \widetilde p \over 1 + 2 \varphi}
       \left( {\widehat\gamma^2 \widehat v^i \widehat v_i \over c^2}
       + {1 \over a {\cal N}} \chi^i {\widehat \gamma^2 \widehat v_i \over c}
       \right), \quad
       \widetilde T^0_i
       = {1 \over {\cal N}} \left( \widetilde \mu + \widetilde p \right)
      {\widehat\gamma^2 \widehat v_i \over c} ,
   \nonumber \\
   & &
       \widetilde T_{ij}
       = a^2 \left[ \left( 1 + 2 \varphi \right) \widetilde p g^{(3)}_{ij}
       + \left( \widetilde \mu + \widetilde p \right) {\widehat\gamma^2 \widehat v_i \widehat v_j \over c^2} \right].
   \label{Tab-pert}
\eea
The ADM fluid quantities are
 \bea
   & & E = \widetilde \mu
       + \left( \widetilde \mu + \widetilde p \right) \left( \widehat\gamma^2 -1 \right), \quad
       J_i = a \left( \widetilde \mu + \widetilde p \right) {\widehat\gamma^2 \widehat v_{i} \over c}
       , \quad
       J^i = {(\widetilde \mu + \widetilde p ) \over a (1 + 2 \varphi)}
       {\widehat\gamma^2 \widehat v^{i}\over c}
       ,
   \nonumber \\
   & & S^i_j
       = \widetilde p \delta^i_j
       + {\left( \widetilde \mu + \widetilde p \right) \over (1+2\varphi)}
       {\widehat\gamma^2 \widehat v^{i}\widehat v_{j} \over c^2 }, \quad
       S = 3 \widetilde p + (\widetilde \mu + \widetilde p) (\widehat\gamma^2 -1),
       \nonumber \\
   & &    \overline{S}^i_j
       = (\widetilde \mu + \widetilde p )
       \left[ {\widehat\gamma^2\over 1+2\varphi } {\widehat v^i\widehat v_j \over c^2}
                          - {1 \over 3} \delta^i_j (\widehat \gamma^2 -1 ) \right].
\eea

%%%%%%%%%%%%%%%%%%%%%%%%%%%%%%%%%%%%%%%%%%%%%%%%%%%%%%%%%%%%%%%

\section*{References}
\begin{thebibliography}{10}
%\bibitem{Einstein-1917}
%         A. Einstein, K\"oniglich Preussische Akademie der Wissenschaften (1917);
%         translated in J. Bernstein and G. Feinberg, eds,
%         {\it Cosmological Constants: Papers in Modern Cosmology} (Columbia Univ. Press, 1989).
\bibitem{Bardeen-1988}
         J.M. Bardeen, Particle Physics and Cosmology,
         edited by L. Fang and A. Zee (Gordon and Breach, London, 1988).
\bibitem{Hwang-1991}
         J. Hwang, Atsrophys. J. \textbf{375}, 443 (1991).
%\bibitem{Hwang-1993}
%         Hwang, J. 1993, Phys. Rev. D, 48, 3544
\bibitem{Hwang-2001}
         J. Hwang and H. Noh, Phys. Rev. D, \textbf{65}, 023512 (2001).
\bibitem{Third-order-2005}
         J. Hwang and H. Noh, Phys. Rev. D, \textbf{72}, 044012 (2005).  
\bibitem{NL-2004}
         H. Noh and J. Hwang, Phys. Rev. D, \textbf{69}, 104011 (2004).    
\bibitem{Hwang-2013}
         J. Hwang and H. Noh, Mon. Not. R. Astron. Soc. \textbf{433}, 3472 (2013). 
\bibitem{York-1973}
         J.W. York, J. Math. Phys., \textbf{14}, 456 (1973).          
\bibitem{Ehlers-1993}
         J. Ehlers, Proceedings of the mathematical-natural science of
         the Mainz academy of science and literature, Nr. \textbf{11}, 792 (1961);
         English translation, Gen. Rel. Grav. \textbf{25}, 1225 (1993).
\bibitem{Ellis-1971}
         G.F.R. Ellis, General relativity and cosmology, Proceedings of
         the international summer school of physics Enrico Fermi course 47,
         edited by R.K. Sachs, (Academic Press, New York, 1971).
\bibitem{Ellis-1973}
         G.F.R. Ellis, Cargese Lectures in Physics, edited by
         E. Schatzmann, (Gorden and Breach, New York, 1973).
%\bibitem{Ellis-1984}
%         Ellis, G. F. R. 1984, General Relativity and Gravitation, edited by
%                       Bertotti, B. et al. (Reidel, Dordrecht)
%\bibitem{Ellis-Stoeger-1987}
%         Ellis G. F. R. \& Stoeger, W. 1987, Class. Quant. Grav. 4, 1697
%\bibitem{Friedmann-1922}
%         A.A. Friedmann, Zeitschrift f\"ur Physik, \textbf{10}, 377 (1922);
%         English translation (1999), Gen. Rel. Grav. \textbf{31}, (1991).
%\bibitem{Harrison-1967}
%         Harrison, E. R. 1967, Rev. Mod. Phys., 39, 862
\bibitem{Hawking-Ellis-1973}
         S.W. Hawking and G.F.R. Ellis, The large scale structure of space-time, (Cambridge University Press, Cambridge, 1973).                            
\bibitem{ADM}
         R. Arnowitt, S. Deser and C.W. Misner, Gravitation: an introduction to current research, edited by L. Witten (Wiley, New York, 1962).
%\bibitem{Bardeen-1980}
%         Bardeen, J. M. 1980, Phys. Rev. D, 22, 1882
%\bibitem{Clarson_etal-2011}
%         Clarkson, C., Ellis, G., Larena, J. \& Umeh, O. 2011, Rep. Prog. Phys. 74, 112901
\bibitem{Hwang-2006}
         J. Hwang and H. Noh, Phys. Rev. D, \textbf{73}, 044021 (2006).
\bibitem{NL-2007}
         J. Hwang and H. Noh, Phys. Rev. D, \textbf{76}, 103527 (2007).
%\bibitem{HNG-2012}
%         Hwang, J., Noh, H. \& Gong, J. 2012, ApJ, in press
%\bibitem{HNP-2010}
%         Hwang, J., Noh, H. \& Park, C-G. 2010, Phys. Rev. D, 82, 043525
%\bibitem{JGNH-2011}
%         Jeong, D., Gong, J., Noh H. \& Hwang, J. 2011, ApJ, 727, 1
%\bibitem{KS-1984}
%         Kodama, H. \& Sasaki, M. 1984, Prog. Theor. Phys. Suppl., 78, 1
%\bibitem{Lifshitz-1946}
%         E.M. Lifshitz, J. Phys. (USSR), \textbf{10}, 116 (1946).
%\bibitem{Lyth-Malik-Sasaki-2005}
%         Lyth, D. H., Malik, K. A. \& Sasaki, M. 2005, JCAP, 05, 004
%\bibitem{Ma-Bertschinger-1995}
%         Ma, C. \& Bertschinger, E. 1995, ApJ, 455, 7
%\bibitem{Mukhanov-1988}
%         Mukhanov, V. F. 1988, Sov. Phys. JETP, 68, 1297
%\bibitem{MFB-1992}
%         Mukhanov, V. F., Feldman, H. A. \& Brandenberger, R. H. 1992,
%                        Phys. Rep., 215, 203
%\bibitem{Nariai-1969}
%         Nariai, H. 1969, Prog. Theor. Phys., 41, 686
%\bibitem{Peebles-1980}
%         P.J.E. Peebles, The Large-Scale Structure of the Universe
%         (Princeton University Press, Princeton, 1980)
%\bibitem{Sachs-Wolfe-1967}
%         Sachs, R. K. \& Wolfe, A. M. 1967, ApJ, 147, 73
%\bibitem{Sasak-1986}
%         Sasaki, M. 1986, Prog. Theor. Phys. 76, 1036
%\bibitem{Smarr-York-1978}
%         Smarr, L. \& York, J. W. 1978, Phys. Rev. D, 17, 2529
%\bibitem{Vishniac-1983}
%         E.T. Vishniac, Mon. Not. R. Astron. Soc. \textbf{203}, 345 (1983).
%\bibitem{ZN-1983}
%         Zel'dovich Ya. B. \& Novikov, I. D. 1983, Relativistic astrophysics,
%         Vol 2, The structure and evolution of the universe,
%         (Univ. Chicago Press, Chicago)








\end{thebibliography}
\end{document}